\documentclass{article}
\usepackage{spconf,amsmath,graphicx,hyperref}
\usepackage{tabularray}
\usepackage{float}
\usepackage{xcolor}
\usepackage[numbers,sort&compress]{natbib}

\title{Teaching Machines to Speak using Articulatory Control}
\name{
  Akshay Anand, Chenxu Guo, Cheol Jun Cho$^{*}$, Jiachen Lian$^{*\dagger}$, Gopala Anumanchipalli$^{*}$%
  \thanks{$^{*}$ Equal advising, $^{\dagger}$ Project lead, jiachenlian@berkeley.edu}
}
\address{University of California, Berkeley}

\begin{document}
\ninept

\maketitle

\vspace{-3mm}

\begin{abstract}
Current speech production systems predominantly rely on large transformer models that operate as black boxes, providing little interpretability or grounding in the physical mechanisms of human speech. We address this limitation by proposing a new framework: speech generation through explicit articulatory control. This reframes speech as a motor control task similar to robotic manipulation. Our approach uses reinforcement learning to train a policy that directly controls the movements of vocal tract articulators, such as the tongue, lips, and jaw, to produce syllable-level speech. Specifically, we employ the Proximal Policy Optimization algorithm to learn optimal articulatory movements based on acoustic feedback provided by our audio perceiver, Sylber. The resulting articulatory trajectories are decoded into audio using SPARC, a pre-trained articulatory-to-speech decoder. We train this framework on six target syllables, and it demonstrates successful convergence, with similarity scores between the policy-generated audio and the target syllables exceeding 0.85. Accurate human transcription of the audio for syllables such as “please”, “loot”, and "cat" demonstrates the intelligibility of this framework.
\end{abstract}
\vspace{-3mm}

\section{Introduction}
\label{sec:intro}
Over the past decade, the field of speech synthesis has been transformed by the advent of large generative models~\cite{zhang2023survey}. Trained on massive audio corpora, these systems are capable of producing natural and expressive speech by learning complex mappings from input signals to intermediate acoustic embeddings. The naturalness and expressiveness of such systems scale with both data size and model capacity. However, they still exhibit notable limitations in flexibility—namely, the ability to adjust speech through fine-grained controls—and in explainability. The latter is particularly critical in speech healthcare applications~\cite{lian2024ssdm, ruiter2025challenges}, where interpretability provides a pathway toward clinically reliable diagnosis of motor speech disorders and related conditions.

In contrast, human speech production is physically grounded and inherently interpretable~\cite{fant1971acoustic}. Speech emerges from the synergy and coordination of articulatory movements constrained by the laws of biomechanics, offering a natural foundation for both flexibility and explainability~\cite{browman1992articulatory}. This raises a central question: \textit{can speech generation be modeled in a manner that more closely reflects how humans actually produce speech?}

Human speech arises from a highly coordinated process of motor control~\cite{browman1995dynamics, tourville2011diva}. Vocal tract articulators such as the tongue, lips, and jaw are precisely coordinated to shape airflow and generate sound~\cite{rebernik2021review}. This process is dynamic, interpretable, and biomechanically grounded: each sound corresponds to a specific configuration and trajectory of articulators~\cite{chiba1958vowel, fant1971acoustic, maeda1990compensatory, international1999handbook}. Crucially, speech production is also shaped by multiple feedback mechanisms, including auditory feedback, somatosensory feedback, and proprioceptive feedback, which allow speakers to monitor and adjust their articulation in real time~\cite{houde2011speech, purcell2006compensation, houde1998sensorimotor, golfinopoulos2011fmri, lametti2012sensory}. Building on this background, one can ask \textit{whether speech generation can be modeled explicitly as a feedback control system~\cite{sastry2013nonlinear} that governs articulatory movements to produce speech}.

Early work developed purely white-box, modularized models that build on articulatory dynamics~\cite{ramanarayanan2016new-dynamics1, hirayama1993inverse-dynamics2}. However, rule-based expert systems limit their generalization ability. Recent efforts have begun to bridge this gap by introducing neural articulatory representations. For example, the Speech Articulatory Coding (SPARC) framework~\cite{cho2024coding} encodes speech into vocal tract kinematics (positional trajectories of articulators) along with source features such as pitch and loudness. SPARC establishes a promising link between articulatory dynamics and acoustic outcomes. However, SPARC does not generate speech in the same way humans produce it. Another line of work attempts to learn articulatory dynamics—termed neural gestural scores~\cite{lian2022deep-gesture1, lian2023articulatory-gesture2}—from kinematic data. Nevertheless, it remains unclear how speech production emerges through dynamic control systems.

In this paper, we propose an articulatory control–based framework for speech production that aligns with how humans actually speak. Instead of predicting acoustics directly, our model generates speech by explicitly controlling the movements of articulators over time. In this way, it learns to emulate human speech production. Unlike many purely neural network–based approaches, our pipeline is fully interpretable, enabling the explanation of subtle speech patterns. We synthesized six fundamental syllables and evaluated their intelligibility through human perception tests. The high recognition accuracy provides strong evidence for the effectiveness of modeling speech production at the syllable level. Although our model does not yet match the performance of current data-driven neural speech synthesis systems, \textit{it demonstrates the feasibility of developing a white-box speech generation model that is fully verifiable and paves the way for future improvements in flexibility, security and clinical interpretability}.
\vspace{-3mm}
\section{Speech Production through Muscle Control}
\label{sec:format}


We frame speech generation as a feedback control problem~\cite{sastry2013nonlinear}, where the task is to determine how articulators move over time to produce a desired sound. This is directly analogous to robotic control: just as a robot with \(N\) joints learns policies to move its actuators toward a goal, a “robotic mouth” must learn policies to coordinate the tongue, lips, and jaw to generate speech. 

As a first step, we focus on generating speech at the syllable level. Syllables are the linguistically defined unit of speech production~\cite{macneilage1998frame-syllable, cho2024sylber}: they are small enough to make the learning problem feasible while still requiring meaningful coordination of the mouth. By contrast, attempting to learn full sentence generation from scratch would require lots of exploration across a very large number of episodes.

Finally, we draw inspiration from how humans acquire speech. Infants do not learn to speak by memorizing mouth movements; rather, they learn through trial and error, gradually refining their motor control of the mouth using acoustic feedback~\cite{kuhl2004early-infants, goldstein2003social, goldstein2008social, eilers1994infant}. To emulate this process, we employ an online reinforcement learning approach in which the policy iteratively improves by attempting to produce target syllables, receiving feedback, and adjusting accordingly.

\begin{figure}
  \centering
  \includegraphics[width=0.45\textwidth]{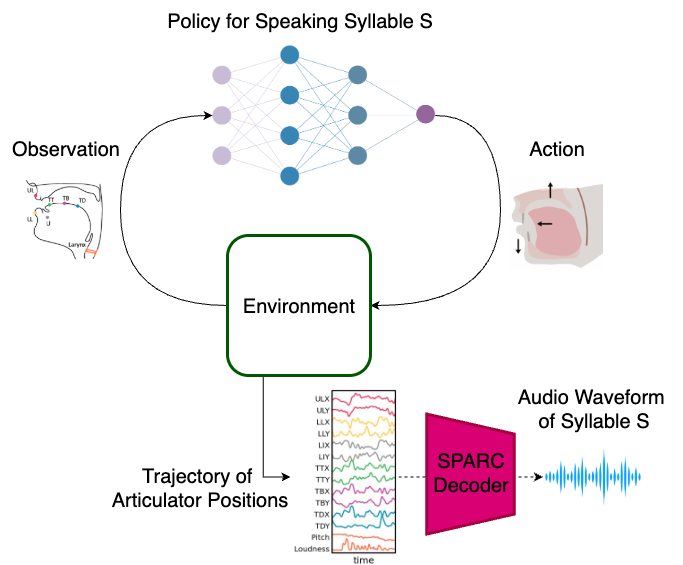}
  \caption{{\bf Process for Speaking a Syllable.} For some syllable \(S\), our policy continuously receives an observation from the environment and takes an action, and at the end, we fetch the generated trajectory of positions and decode it into audio.}
  \label{fig:1}
\end{figure}

\subsection{Environment Design}
\label{ssec:subhead2}

\begin{figure}
  \centering
  \includegraphics[width=0.5\textwidth]{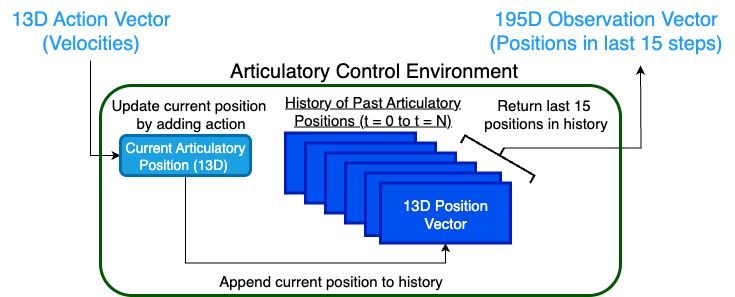}
  \caption{{\bf Articulator-based Environment} We show the process of how our environment processes actions and updates our state to return an observation.}
  \label{fig:2}
\end{figure}

Our environment is designed to simulate the dynamics of articulatory control while remaining amenable to reinforcement learning. The agent interacts with this environment by controlling a set of articulators and receiving observations that describe the current and previous states of the system. A schematic overview of this environment design is shown in Figure~\ref{fig:2}.

The agent directly controls six articulators~\cite{cho2024coding}: the tongue dorsum (TD), tongue blade (TB), tongue tip (TT), lower incisor (LI), upper lip (UL), and lower lip (LL). Each articulator moves along two spatial axes (X and Y), resulting in 12 controllable degrees of freedom. In addition, the agent modulates vocal loudness (L), yielding a total of 13 continuous control dimensions. At each timestep \(t\), the action \(\mathbf{a}_t\) specifies how much to move each articulator and how to adjust loudness, effectively representing articulatory velocities. 

\[
\mathbf{a}_t =
\begin{bmatrix}
(V_x, V_y)^{\text{TD}} \\
(V_x, V_y)^{\text{TB}} \\
(V_x, V_y)^{\text{TT}} \\
(V_x, V_y)^{\text{LI}} \\
(V_x, V_y)^{\text{UL}} \\
(V_x, V_y)^{\text{LL}} \\
V^{\text{L}}
\end{bmatrix}
\in [-0.5,\, 0.5]^{13}
\]

Our action is the velocity ($v$) we apply, so it would make sense for the state \(s\) to be the current position. However, a single snapshot of position is not enough information because it doesn't show the direction of movement. For example, an object could be at the same location but moving upward or downward. To solve this problem of partial observability, we use frame stacking. We define the state ($s$) as the last 15 frames of $X$, $Y$ positions for each articulator, and loudness values. This technique gives the system a short-term memory of its recent trajectory, which helps it produce smooth and coordinated movements.

\[
\mathbf{s}_t =
\begin{bmatrix}
(x,y)^{\text{TD}}_{t-14}, \;\dots,\; (x,y)^{\text{LL}}_{t-14}, \; L_{t-14} \\
\vdots \\
(x,y)^{\text{TD}}_{t}, \;\dots,\; (x,y)^{\text{LL}}_{t}, \; L_{t}
\end{bmatrix}
\in 
 [-3,3]^{13 \cdot 15}
\]

At the start of each episode, the environment is configured with a target syllable (in the form of an embedding), which represents the sound we want to speak. The agent generates articulatory movements step by step, which are tracked in one trajectory. This trajectory are subsequently mapped to audio using SPARC’s decoder, and feedback is provided to the agent regarding how well the generated output matches the target.

\subsection{Acoustic Feedback}
\label{ssec:subhead5}
In our RL framework, the reward serves as the feedback signal that guides the policy to improve. The central challenge is designing feedback that meaningfully reflects how closely the policy-generated speech matches the intended target syllable.

Sylber~\cite{cho2024sylber} is a framework that creates embeddings for syllables directly in speech audio. Unlike phoneme- or frame-level representations, syllable embeddings capture information over longer temporal windows, reflecting the natural organization of speech into syllables. Sylber not only provides an embedding representation of each syllable but also includes an automatic detection mechanism that identifies syllable boundaries in speech and associates each detected unit with its learned embedding. In our work, Sylber is valuable both as a tool for obtaining meaningful representations of syllables and as a perception model to analyze the speech produced by our policy.

\[
reward_t = \frac{\text{SylberEmb}^{\text{policy}}_t \cdot \text{SylberEmb}^{\text{target}}_t}
       {\|\text{SylberEmb}^{\text{policy}}_t\| \, \|\text{SylberEmb}^{\text{target}}_t\|}, 
\quad t = 1, \dots, T
\]

\begin{figure}
  \centering
  \includegraphics[width=0.45\textwidth]{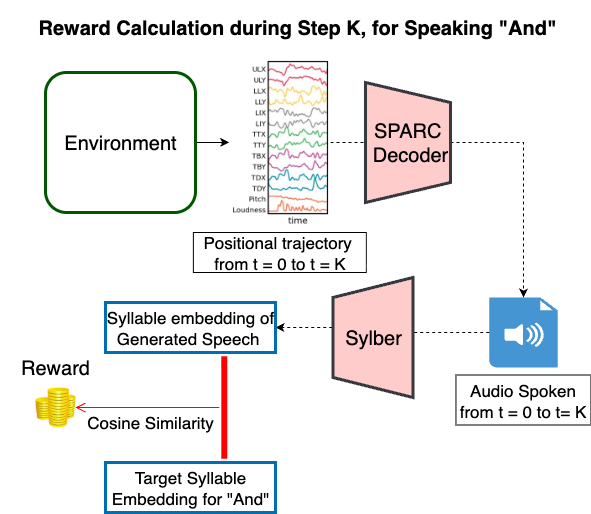}
  \caption{{\bf Sylber Reward Calculation} This shows the process of calculating the reward at timestep K based on the information stored in the environment. We convert the positional trajectory so far into an audio waveform and then extract detected syllable embeddings from it and compare them to the target syllable.}
  \label{fig:3}
\end{figure}

In our setup, the articulatory trajectory produced by the policy is decoded into audio waveform using SPARC’s decoder. The resulting waveform is then processed by Sylber, which detects the syllables being produced and outputs their corresponding embeddings. At each timestep, we focus on the embedding of the most recently detected syllable and compare it to the embedding of the target syllable using cosine similarity. This comparison is performed step-by-step throughout the episode: after every frame of articulatory movement, the current partial trajectory is decoded into audio, and similarity is computed. The reward provided to the agent at each step is this similarity score \([-1,1]\), with higher values indicating that the generated speech more closely resembles the intended syllable (Figure~\ref{fig:3}).

If Sylber fails to detect any syllable in the generated audio at a given step, we assign a negative reward of \(-1\). This penalty discourages the agent from producing unstructured or unintelligible articulatory movements, reinforcing the importance of generating acoustically valid syllables. 

\subsection{Choice of Reinforcement Learning Algorithm}
\label{ssec:subhead3}

Our objective is to train an agent in a manner that parallels how humans learn speech. Human speech learning is fundamentally interactive: people refine their vocal control by producing sounds, listening to their resulting sound, and iteratively adjusting articulator movements based on feedback from their parents. This motivates us to adopt an online reinforcement learning framework, where the agent improves its policy directly through trial-and-error interaction with the environment, rather than relying solely on pre-collected datasets. 

We narrow down to using an on-policy algorithm. In an on-policy setting, the policy is updated using data generated by its own actions. This characteristic is especially important in our domain: articulators demand fine-grained control, so the training data distribution must remain closely aligned with the policy’s current exploration behavior.  Among on-policy methods, we select \textit{Proximal Policy Optimization (PPO)} as the learning algorithm~\cite{schulman2017proximal-ppo}. PPO has emerged as a standard in reinforcement learning due to its balance between stability and efficiency. 

\section{Experiments and Results}
\label{sec:typestyle}
We use this reinforcement learning framework to train separate policies for 6 different syllables: \textit{please}, \textit{road}, \textit{fan}, \textit{loot}, \textit{cat}, and \textit{age}. This set of syllables includes stop consonants (/p/, /t/), fricatives (/s/, /z/, /f/), nasals (/n/), laterals (/l/), high and low vowels (/iy/, /æ/), and diphthongs (/ey/). By covering a range of phonemes, as well as both consonantal and vocalic gestures, these syllables provide a benchmark for testing whether our control-based model can generalize across different classes of speech sounds.

We train our articulatory control policy (for each syllable) with PPO over 25,000+ episodes, each lasting exactly 50 timesteps (approximately one second of speech). At the start of each episode, all articulators are reset to a position of zero. At each step, the policy determines the optimal action from the observation of the environment. We take this action, then the environment provides Sylber-based acoustic feedback, and PPO updates the actor-critic networks using the clipped surrogate loss. Both actor and critic are multi-layer perceptrons (MLPs). 

To encourage exploration in this high-dimensional, continuous action space, we initialize the action distribution with a high standard deviation (0.7), which is gradually decayed by 0.01 every 100 episodes until reaching 0.05. This schedule allows broad early exploration of articulatory trajectories before converging to stable, controlled speech production.  
\vspace{-3mm}

\subsection{Rewards and Similarity Scores}
\label{ssec:subhead3}

\begin{figure}
  \centering
  \includegraphics[width=0.34\textwidth]{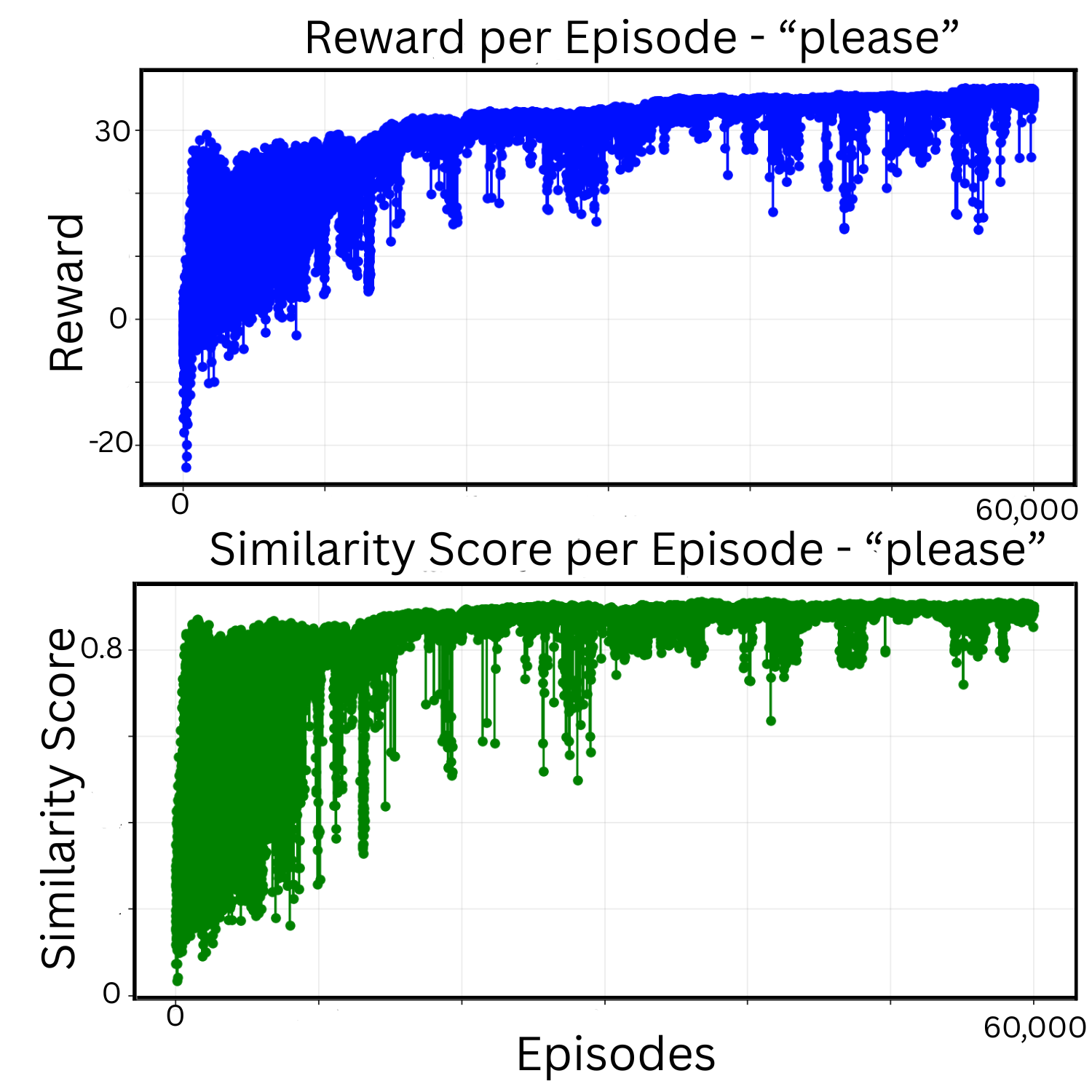}
  \caption{{\bf Reward Graphs} These plots show the trend of reward and similarity score as training progresses for the syllable "please".}
  \label{fig:4}
\end{figure}

We track reward curves to evaluate policy improvement over training. Since PPO relies on stable learning, we look for an overall upward trend rather than erratic spikes. As shown in Figure~\ref{fig:4}, rewards increase consistently for $\pi_{\text{please}}$, and although not shown, the same pattern is evident for the other syllables' policies. Early training is marked by high variability and negative spikes due to exploration, but the curves stabilize as the policies converge to reliable articulatory strategies. We also track the highest similarity score per episode, indicating how closely the policy is to speaking the target syllable. These steadily improve across episodes, having a similar trend to the reward. This confirms that the reward is well shaped and correlated by similarity to the target.

\subsection{Articulatory Trajectories}
\label{ssec:subhead3}
\begin{figure}
  \centering
  \hspace{-0.5cm}
  \includegraphics[width=0.48\textwidth]{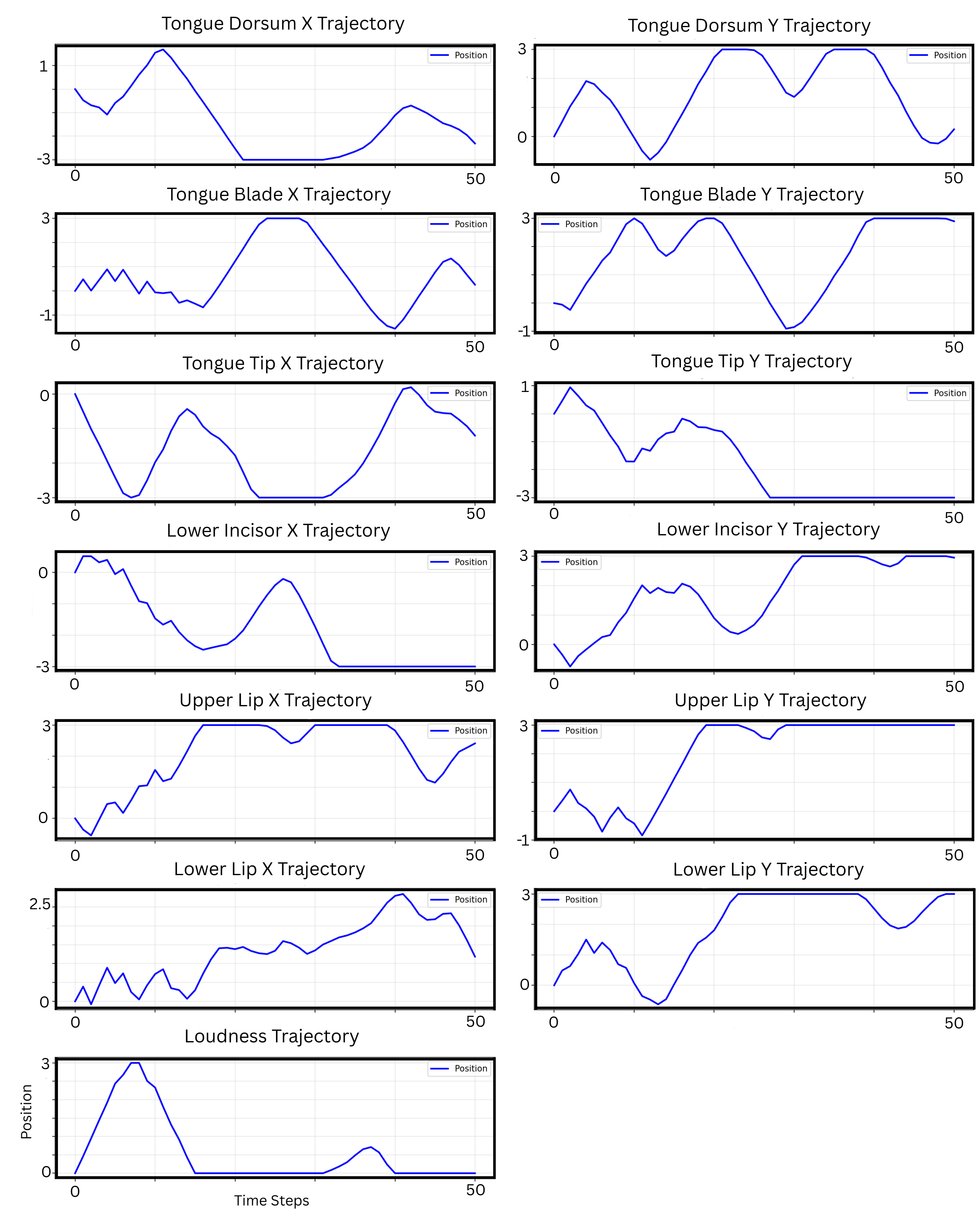}
  \caption{{\bf Articulatory Trajectory of "Please"} These plots show the position vs time for the X,Y direction for all 6 articulators, as well as the loudness level vs time. This trajectory is generated by the policy trained to speak "please"}
  \label{fig:5}
\end{figure}

We evaluate the policy by running a full episode of $\pi_{\text{please}}$ after 60{,}000 training episodes, tracking the movements of the tongue dorsum and blade, tongue tip, lips, and lower incisors along with loudness over 50 timesteps (Figure~\ref{fig:5}). The loudness curve highlights the syllable boundaries: the main peak from steps 0--18 corresponds to the production of \textit{please}, while a smaller later peak reflects auxiliary noise. Within this main interval, the articulatory dynamics follow the expected sequence /p l iy z/.  

During /p/ (steps 0--7), the lower lip rises while the upper lip lowers slightly, creating a bilabial closure; loudness remains near zero until the burst release. Between steps 7--12, the tongue tip elevates toward the alveolar ridge to form the /l/ constriction, while the tongue dorsum lowers but also advances forward due to coarticulation with /iy/. The vowel /iy/ (steps 12--15) is marked by dorsum elevation that establishes the high front posture. Finally, /z/ overlaps weakly with /iy/ between steps 12--18: the tongue tip and blade rise, but loudness decays rapidly, leaving truncated frication.   Overall, these trajectories show that the policy not only maximizes reward but also reproduces interpretable, phonetically consistent motor patterns for each phoneme in \textit{please}.

\subsection{Human Evaluation}
\label{ssec:subhead3}
Once our policy has converged, we evaluate the intelligibility of the speech by having a human reviewer transcribe it. The results are shown in the table below. For \textit{cat}, \textit{loot}, and \textit{please}, our generated speech is transcribed correctly. However, for \textit{fan} and \textit{road}, we transcribe them as \textit{roar} and \textit{fang}, showing that the policy learns the first part of the syllable but doesn't correctly get the ending. Our policy is not able to produce \textit{age} well, as it being transcribed as \textit{we}.


\begin{table}[h!]
\centering
\caption{Contains the total rewards, similarity score to target syllable, and human transcription of generated audio for each syllable's policy after training.}
\label{tab:syllable_scores}
\begin{tblr}{
  column{even} = {c},
  column{3} = {c},
  column{5} = {c},
  column{7} = {c},
  hline{1-2,5} = {-}{},
}
Syllable            & please & road  & fan   & loot  & cat   & age   \\
Reward        & 37.41  & 32.89 & 28.26 & 33.12 & 32.31 & 30.87 \\
Similarity    & 0.92   & 0.87  & 0.79  & 0.85  & 0.89  & 0.92  \\
Human & please & roar  & fand  & loot  & cat   & we    
\end{tblr}
\end{table}

\vspace{-3mm}

\section{Discussion}
\label{sec:concl}

In this work, we present an articulatory reinforcement learning framework for syllable-level speech generation, shifting from black-box generative models to control theory. Instead of large transformer architectures, lightweight multilayer perceptrons (MLPs) determine articulator movements from current positions and short-term history.

A central strength of this approach is interpretability: motor commands for each articulator are explicitly modeled, allowing trajectories to be traced back to phonetic targets and providing insight into control strategies. By training policies from scratch, the framework also parallels how infants acquire speech—exploring articulatory space and refining motor control through feedback.

This proof-of-concept demonstrates that reinforcement learning can produce intelligible, interpretable speech and offers a computational lens on speech development. Although the current model does not yet achieve the raw perceptual quality of large data-driven neural synthesis systems, it establishes the feasibility of a white-box TTS framework—one that is auditable, verifiable, and amenable to clinical interpretability~\cite{lian2023unconstrained-udm, lian-anumanchipalli-2024-towards-hudm, ssdm, lian2024ssdm2.0, zhou2024yolostutterendtoendregionwisespeech, zhou2024stutter, zhou2024timetokensbenchmarkingendtoend, zhou2025phonetic-error-detection, guo2025dysfluentwfstframeworkzeroshot, zhang2025analysisevaluationsyntheticdata, ye2025lcs0ctc, ye2025seamlessalignment-neurallcs, li2025k}. Looking ahead, goal-conditioned reinforcement learning can enable a single policy to generate diverse syllables and naturally scale toward sentence-level production. Overall, this framework provides an efficient, transparent, and human-like alternative to conventional speech synthesis, while laying the groundwork for trustworthy and secure deployment in sensitive healthcare and educational contexts. 

\section{Acknowledgements}
We are deeply grateful to Baifeng Shi (UC Berkeley) for the insightful discussions and invaluable perspectives.

\bibliographystyle{IEEEbib}
\bibliography{strings,refs}

\end{document}